%% file: majun.tex
\documentstyle[aaspp4,psfig]{article}

\newcommand\beq{\begin{equation}} 
\newcommand\eeq{\end{equation}}

\received{} 
\accepted{} 

\lefthead{Ma et al.} 
\righthead{Spectral Energy and Age Distributions for 51 Globular Cluster Candidates}

\begin{document}

{\title{Spectral Energy and Age Distributions for 51 Globular Cluster Candidates}

\author{ 
Jun Ma\altaffilmark{1}, 
Xu Zhou\altaffilmark{1},
Jiansheng Chen\altaffilmark{1},
Hong Wu\altaffilmark{1},
Xu Kong\altaffilmark{2},
Zhaoji Jiang\altaffilmark{1}, 
Jin Zhu\altaffilmark{1},
and
Suijian Xue\altaffilmark{1}
}

\altaffiltext{1}{National Astronomical Observatories, 
Chinese Academy of Sciences, Beijing, 100012, P. R. China;
majun@vega.bac.pku.edu.cn}

\altaffiltext{2}{Center for Astrophysics, University of Science and
        Technology of China, Hefei, 230026, P. R. China}

\authoremail{majun@vega.bac.pku.edu.cn}

\begin{abstract}

This paper is the fourth in a series presenting
spectrophotometry of 51 globular cluster candidates,
that were detected by Mochejska et al. in the nearby
galaxy M33 using the data collected by the DIRECT project.
The frames of M33 in this study were taken as part
of the BATC Multicolor Sky Survey. We obtained the spectral
energy distributions of these candidates in 13 intermediate-band
filters. By comparing the integrated photometric measurements with
theoretical stellar population synthesis models of Bruzual \&
Charlot, we estimated
their ages. The BC96 models provide the evolution in time
of the spectrophotometric properties of simple stellar populations
for a wide range of stellar metallicity.
Our results show that half of the candidates are younger than
$10^8$ years, whose age degeneracy is not pronounced.
We also find that globular clusters formed continuously in M33 from
$\sim 4\times10^6$ -- $10^{10}$ years. Our results are in agreement
with Chandar et al., who estimated ages for 35 globular clusters
candidates in common by comparing the photometric measurements to
integrated colors from theoretical models by Bertelli et al.
The Kolmogorov-Smirnov Test shows that the maximum value of the
absolute difference of estimated ages between Chandar et al. and
us is 0.48, and the significance level probability is 100.00 per cent.
\end{abstract}


\section{INTRODUCTION}

The Galactic globular clusters, which are thought to be among
the oldest radiant objects in the universe and can be
accurately dated in the Galaxy, provide vitally important
information regarding the minimum age of the universe
and the early formation history of our Galaxy.
The study of these systems has contributed much to our
knowledge of stellar evolution and galactic structure.
However, it is necessary to make sure that the conclusions
drawn from the study of the Milky Way's globular cluster
system are not biased either because they are somehow
unusual or because our location in the Milky Way
prevents us from fully characterizing its properties.
So, the study of globular clusters in other galaxies
is valuable, especially in Local Group galaxies. 
Besides the Milky Way, a few of other Local Group
galaxies contain globular clusters, such as the
Sagittarius dwarf spheroidal galaxy, the Large
and Small Magellanic Clouds, M31, and M33.
Christian \& Schommer (1982, 1988) cataloged
more than 250 non-stellar sources by visually examining
$14\times 14$ $\rm inch^2$ unfiltered, unbaked, IIa-O
focus plate of M33, and obtained ground based $BVI$
photometry for $\sim 106$ star cluster candidates,
and estimated that M33 contains only $\sim 20$
total ``true'' globular clusters.
Using the $Hubble$ $Space$ $Telescope$ WFPC2 multiband images
of 55 fields in M33, Chandar, Bianchi,
\& Ford (1999a, 2001) detected 162 star clusters,
131 of which were previously unknown. They estimated
the total number of globular clusters in M33 to be
$75\pm 14$. Especially, Mochejska et al. (1998)
detected 51 globular cluster candidates in M33, 32 of
which were not previously cataloged, using the data
collected in the DIRECT project ({\cite{Kaluzny98}; {\cite{Stanek98}).

M33 is a small Scd Local Group galaxy, about 15 times farther from
us than the LMC (distance modulus is 24.64) (Freedman,
Wilson, \& Madore 1991; Chandar, Bianchi, \&
Ford 1999a). It is interesting and important because
it represents a morphological type intermediate
between the largest ``early-type'' spirals and the dwarf
irregulars in the Local Group (Chandar, Bianchi, \& Ford 1999a). 
At a distance of $\sim 840$ kpc, M33 is the only
nearby late-type spiral galaxy, it can provide an
important link between the cluster populations of earlier-type
spirals (Milky Way galaxy and M31) and the numerous,
nearby later-type dwarf galaxies.
Our collaboration, the Beijing-Arizona-Taiwan-Connecticut
(BATC) Multicolor Sky Survey (Fan et al. 1997; Zheng et al. 1999),
already had M33 spiral galaxy as part of its galaxy
calibration program.

Sarajedini et al. (1998) selected ten halo globular clusters
from Schommer et al. (1991) by inspecting the difference
between the cluster velocity and the disk velocity as
a function of the integrated cluster color, and
constructed color-magnitude diagrams to estimate the cluster
metallicity using the shape and color of the red giant branch.
Under the assumption that cluster age is the global second
parameter, Sarajedini et al. (1998) presented that the average
age of halo globular clusters in M33 appears to be a few Gyrs younger
than halo clusters in the Milky Way. Ma et al. (2002a)
also estimated their ages by comparing the photometry of
each object with theoretical stellar
population synthesis models for different values of metallicity,
and the results showed that eight clusters
have ``intermediate'' ages, i.e. between 1 and 8 Gyrs.

In this paper, we present the SEDs of 37 globular cluster
candidates that were detected by Mochejska et al (1998) in M33,
and age estimates for these candidates
by comparing the integrated photometric measurements with
theoretical stellar population synthesis models.
The multi-color photometry is powerful
to provide  the accurate SEDs for these
stellar clusters.

The outline of the paper is as follows.  Details of observations
and data reduction are given in section 2. In section 3, we provide
a brief description of the 
stellar population synthesis models of Bruzual \&
Charlot (1996, unpublised, hereafter BC96). The distributions of metallicity and age
are given in section 4.
The summary and discussion are presented in section 5.

\section{SAMPLE, OBSERVATIONS AND DATA REDUCTION}

\subsection{Sample of Globular Cluster Candidates}

The sample in this paper is from Mochejska et al. (1998),
who detected 51 globular cluster candidates using the data
collected in the DIRECT project ({\cite{Kaluzny98}; {\cite{Stanek98}).
In this project, the observations for M33 were done with the
1.2 m telescope at the F. L. Whipple Observatory, using a
thinned, back-side illuminated, AR-coated Loral $2048\times 2048$
CCD. The pixel scale is $0\arcsec{\mbox{}\hspace{-0.15cm}.} 3$.
Mochejska et al. (1998) also presented the photometry for these
candidates using standard Johnson-Cousins $BVI$ filters.
Although searching for star clusters in M33 is sporadical,
this work has been continued (Hiltner 1960; Kron \& Mayall 1960;
Melnick \& D'Odorico 1978; Christian \& Schommer 1982, 1988;
\cite{Mochejska98}; Chandar, Bianchi, \& Ford 1999a, 2001).
19 of the 51 globular cluster candidates in Mochejska et al. (1998)
were previously detected by Melnick \& D'Odorico (1978) or
Christian \& Schommer (1982). Table 2 summarizes the common
clusters in different studies. The clusters with symbol (*)
were also studied by Ma et al. (2001, 2002a, 2002b), who
presented their SEDs in 13 intermediate band filters, and
age estimates by comparing the photometry of each object with theoretical
stellar population synthesis models for different values of
metallicity. In this study, we will present the SEDs and age estimates
for 37 globular cluster candidates that were not found in
Ma et al. (2001, 2002a, 2002b). However, we will plot the age distribution
of the 51 candidates for obtaining the whole picture of these globular
clusters formation.
Figure 1 is the image of M33 in filter BATC07 (5785{\AA}), the circles and
the numbers in which indicate the positions and names of the 51 globular cluster
candidates.

\begin{figure}
\figurenum{1}
\vspace{-5.5cm}
\vspace{18cm}
\caption{The image of M33 in filter BATC07 (5785{\AA}) and the positions
of the sample globular cluster candidates.
The center of the image is located at
${\rm R.A.=01^h33^m50^s{\mbox{}\hspace{-0.13cm}.}58}$,
${\rm decl.=30^\circ39^{\prime}08^{\prime\prime}{\mbox{}\hspace{-0.15cm}.4}}$
(J2000.0). North is up and east is to the left.
}
\label{fig1}
\end{figure} 

\subsection{CCD Image Observation}

The large field multi-color observations of the spiral galaxy M33 were
obtained in the BATC photometric system which has a custom designed
set of 15 intermediate-band filters to do spectrophotometry for
preselected 1 deg$^{2}$ regions of the northern sky. The telescope used is the
60/90 cm f/3 Schmidt Telescope of Beijing Astronomical Observatory (BAO),
located at the Xinglong station. A Ford Aerospace 2048$\times$2048 CCD
camera with 15 $\mu$m pixel size is mounted at the Schmidt focus of the
telescope. The field of view of the CCD is $58^{\prime}$ $\times $ $
58^{\prime}$ with a pixel scale of $1\arcsec{\mbox{}\hspace{-0.15cm}.} 7$.  

The multi-color BATC filter system includes 15 intermediate-band filters,
covering the total optical wavelength range from 3000 to 10000{\AA}
(see Fan et al. 1996). The filters were specifically designed to avoid
contamination from the brightest and most variable night sky emission
lines. A full description of the BAO Schmidt telescope, CCD, data-taking
system, and definition of the BATC filter systems are detailed elsewhere
(\cite{Fan96}; \cite{Zheng99}). The images of M33 covering
the whole optical body of M33 were accumulated in 13 intermediate band
filters with a total exposure time of about 38 hours from September
23, 1995 to August 28, 2000.  The CCD images are centered at ${\rm
R.A.=01^h33^m50^s{\mbox{}\hspace{-0.13cm}.}58}$ and
${\rm decl.=30^\circ39^{\prime}08^{\prime\prime}{\mbox{}\hspace{-0.15cm}.4}}$
(J2000). The dome flat-field images were taken by using a diffuse plate in
front of the correcting plate of the Schmidt telescope. For flux calibration,
the Oke-Gunn primary flux standard stars HD~19445, HD~84937, BD~${+26^{\circ}2606}$,
and BD~${+17^{\circ}4708}$ were observed during photometric nights
(see details from Yan et al. 1999; Zhou et al. 2001). The parameters of
the filters and the statistics of the observations are given in Table 1.

\subsection{Image Data Reduction}

The data were reduced with standard procedures, including bias subtraction
and flat-fielding of the CCD images, with an automatic data reduction
software named PIPELINE I developed for the BATC multi-color sky survey
(\cite{Fan96}; \cite{Zheng99}).
The flat-fielded images of each color were combined
by integer pixel shifting. The cosmic rays and bad pixels were corrected
by comparison of multiple images during combination. The images were
re-centered and position-calibrated using the $HST$ Guide Star Catalog.
The absolute flux of intermediate-band filter images was
calibrated using observations of standard stars. Fluxes as observed
through the BATC filters for the Oke-Gunn stars were derived by convolving
the SEDs of these stars with the measured BATC filter transmission
functions (\cite{Fan96}). {\it Column} 6 in Table 1 gives the zero point
error, in magnitude, for the standard stars in each filter. The formal
errors we obtain for these stars in the 13 BATC filters are $\la 0.02$
mag. This indicates that we can define the standard BATC system to an
accuracy of  $\la 0.02$ mag.

\subsection{Integrated Photometry}

For each globular cluster candidate, the PHOT routine in DAOPHOT
(Stetson 1987) was used to obtain magnitudes.
To avoid contamination from nearby objects, a smaller aperture of
$6\arcsec{\mbox{}\hspace{-0.15cm}.} 8$, which corresponds
to a diameter of 4 pixels in Ford CCDs, was adopted.
Aperture corrections were computed using isolated stars.
The spectral energy distributions (SEDs) in 13 BATC filters for 37
star cluster candidates
were obtained and are listed in Table 3.
For the other 14 candidates, the SEDs can be found
in Ma et al. (2001, 2002a, 2002b).
Table 3 contains the following information: Column (1) is cluster
number that is taken from Mochejska et al. (1998).
Column (2) to Column (14) show the magnitudes in
different bands. Second row for each globular cluster candidate is
the uncertainties of the magnitude in the corresponding bands.
The uncertainties for each filter are taken from by DAOPHOT.

\subsection{Comparison with Previous Photometry}

Using the Landolt standards, Zhou et al. (2002) presented the relationships
between the BATC intermediate-band system and $UBVRI$ broadband system
from the catalogs of Landolt (1983, 1992) and Galad\'\i-Enr\'\i quez,
Trullols, \& Jordi (2000).
We show the coefficients of two relationships
in equations (1) and (2).
\beq
m_B=m_{04}+(0.2218\pm0.033)(m_{03}-m_{05})+0.0741\pm0.033,
\eeq
\beq
m_V=m_{07}+(0.3233\pm0.019)(m_{06}-m_{08})+0.0590\pm0.010.
\eeq
Using equations (1) and (2), we transformed the magnitudes of
51 globular cluster candidates in BATC03, BATC04 and BATC05 bands
to ones in the $B$ band, and in BATC06, BATC07 and BATC08 bands to ones
in $V$ band. For candidates 23 and 30,
we change $m_{05}$ to $m_{04}$ because of the strong emission
in BATC05 band.  Figure 2 plots the comparison of
$V$ (BATC) and ($B$$-$$V$) (BATC) photometry with previously
published measurements (Mochejska et al. 1998).
In this figure, our  magnitudes/colors are on the x-axis, the
difference between our and Mochejska et al. (1998) magnitudes/colors
are on the y-axis. 
In Figure 2, we did not plot the color of candidate 2
because of its large value ($B-V=2.50$) by Mochejska et al. (1998). 
Table 4 also shows this comparison. The mean $V$ magnitude and
color differences (this paper
$-$ the paper (of Mochejska et al. 1998)) are $<\Delta V>
=-0.078\pm0.025$ and $<\Delta (B-V)>=-0.150\pm 0.019$
(not including candidate 2), respectively.
The uncertainties in $B$ (BATC) and $V$ (BATC) have been added linearly,
i.e. $\sigma_B=\sigma_{04}+0.2218(\sigma_{03}+\sigma_{05})$, and
$\sigma_B=\sigma_{07}+0.3233(\sigma_{06}+\sigma_{08})$, to reflect the
error in the three filter measurements.
For the colors, we added the errors in quadrature,
i.e. $\sigma_{(B-V)}={(\sigma_B^2+\sigma_V^2)}^{1/2}$. 
From Figure 2 and Table 4, it can be seen that
there is good agreement in the photometric measurements,
although there exits a error in color.

{\begin{figure}
\figurenum{2}
\centerline{\psfig{file=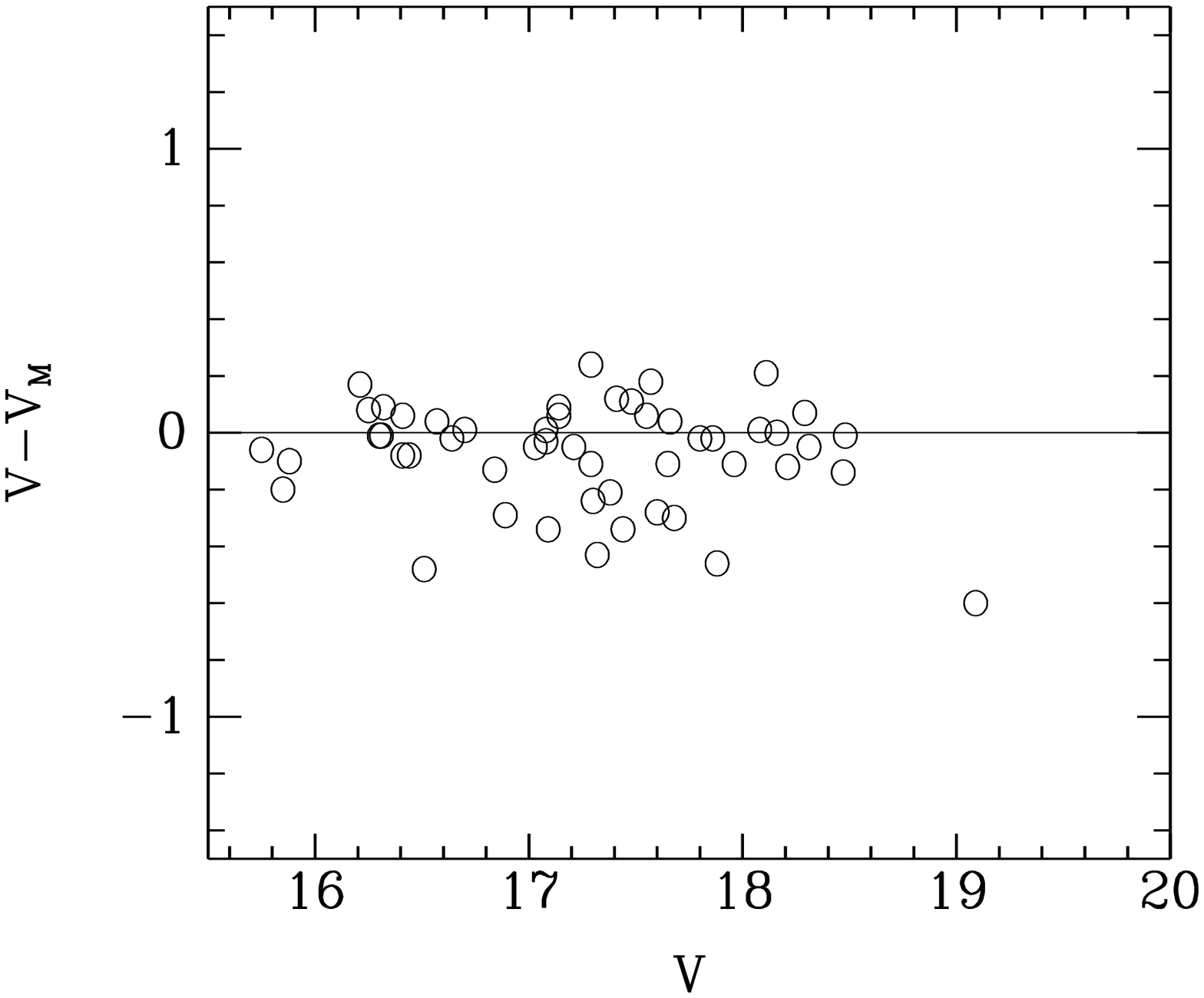,width=16.0cm,angle=-90}}
\label{fig2}
\end{figure}
\begin{figure}
\vspace{-2cm}
\figurenum{2}
\centerline{\psfig{file=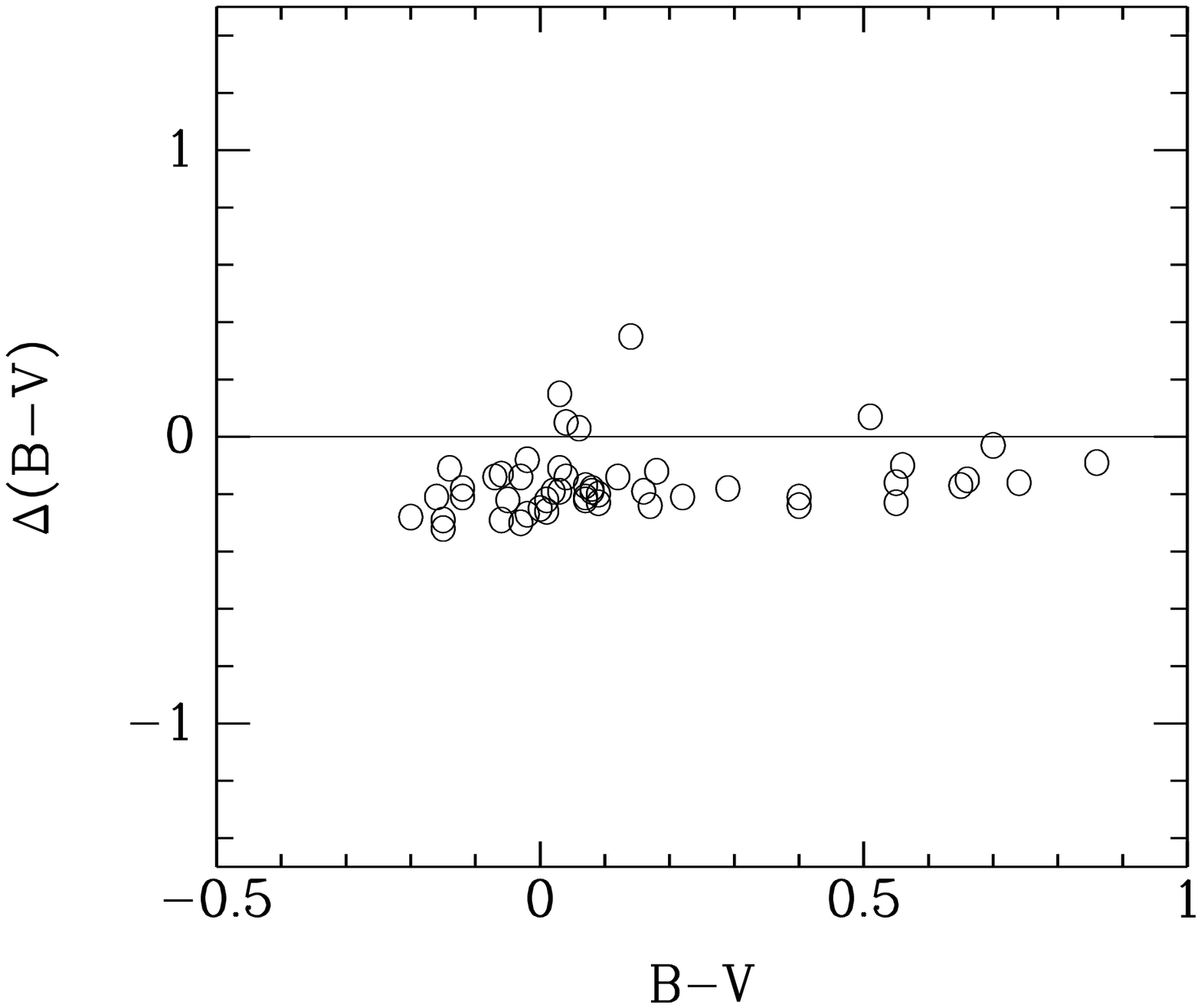,width=16.0cm,angle=-90}}
\vspace{-1cm}
\caption{Comparison of cluster photometry with previous measurements}
\label{fig2}
\end{figure}}

\section{DATABASES OF SIMPLE STELLAR POPULATIONS}

Since the pioneering work of Tinsley (1972) and Searle, Sargent,
\& Bagnuolo (1973),
evolutionary population synthesis has become a standard technique to
study the stellar populations of galaxies. This technique benefits
from the improvement
in the theory of the chemical evolution of galaxies, star formation,
stellar evolution and atmospheres, and of the development of synthesis
algorithms and the availability of various evolutionary synthesis models.
A comprehensive compilation of such models was presented by Leitherer et
al. (1996) and Kennicutt (1998). More widely used models are those from
the Padova and Geneva group (e.g. \cite{Schaerer97}; \cite{Schaerer98};
\cite{Bressan96}; \cite{Chiosi98}), GISSEL96 (\cite{Charlot91};
\cite{Bruzual93}, hereafter BC93; BC96), PEGASE (\cite{Fioc97}) and
STARBURST99 (\cite{Leitherer99}).

A simple stellar population (SSP) is defined as a single generation
of coeval stars with fixed parameters such as metallicity, initial
mass function, etc. (\cite{Buzzoni97}).
SSPs are the basic building blocks of synthetic spectra
of galaxies that can be used to infer the formation and subsequent
evolution of the parent galaxies (\cite{Jab96}).
They are modeled by a collection of stellar evolutionary tracks with
different masses and initial chemical compositions, supplemented
with a library of stellar spectra for stars at different evolutionary
stages in evolution synthesis models.
In order to study the
integrated properties of star clusters in M33, as the first step,
we use the SSPs of Galaxy Isochrone Synthesis Spectra Evolution Library
(hereafter GSSP; BC96) because
they are simple and reasonably well understood.
 
\subsection{SED of GSSPs}

Charlot \& Bruzual (1991) developed a model of
stellar population synthesis. In this model, the
population synthesis method can be used to determine
accurately the distribution of stars in the
theoretical color-magnitude diagram for
any stellar systems. BC93
presented ``isochrone synthesis'' as a natural and
reliable approach to model the evolution of stellar
populations in star clusters and galaxies. With
this isochrone synthesis algorithm, BC93
computed the spectral energy distributions of stellar
populations with solar metallicity. BC96 improved BC93
evolutionary population synthesis models. The updated version
provides the evolution of the spectrophotometric properties for a wide
range of stellar metallicity, which are
$Z=0.0004, 0.004, 0.008, 0.02, 0.05,$ and $0.1$
(see Ma et al. 2001).

\subsection{Integrated Colors of GSSPs}

Using the multi-color photometry,
Kong et al. (2000) have studied the relative chemical abundance,
age, and reddening distributions for different components
of M81. They obtained the best-fit age and reddening
values by minimizing the difference between the observed
colors and the predicted values of the theoretical stellar
population synthesis models of BC96.
To determine the distributions of age for the
globular cluster candidates in this paper, we follow
the method of Kong et al. (2000). Since the observational
data are integrated luminosity, we need to convolve
the SED of GSSP with BATC filter profiles to obtain the optical
and near-infrared integrated luminosity for comparisons (Kong et al. 2000).
The integrated luminosity
$L_{\lambda_i}(t,Z)$ of the $i$th BATC filter can be calculated as
\beq 
L_{\lambda_i}(t,Z) =\frac{\int
F_{\lambda}(t,Z)\varphi_i(\lambda)d\lambda} {\int
\varphi_i(\lambda)d\lambda},
\eeq
where $F_{\lambda}(t,Z)$ is the spectral energy distribution of
the GSSP of metallicity $Z$ at age $t$, $\varphi_i(\lambda)$ is the
response functions of the $i$th filter of the BATC filter system
($i=3, 4, \cdot\cdot\cdot, 15$),
respectively.
To avoid using distance dependent parameters,
we calculate the integrated colors of a GSSP relative to
the BATC filter BATC08 ($\lambda=6075${\AA}):

\beq 
\label{color}
C_{\lambda_i}(t,Z)={L_{\lambda_i}(t,Z)}/{L_{6075}(t,Z)}.  
\eeq

As a result, we obtain intermediate-band colors for 6 metallicities from
$Z=0.0004$ to $Z=0.1$.

\section{RESULTS}

\subsection{Cluster Ages}

Integrated colors of star clusters depend mostly on age,
with a secondary dependence on metallicity.
In order to obtain intrinsic colors of 37 globular cluster
candidates and hence accurate ages,
the photometric measurements must be dereddened.
As Chandar, Bianchi, \& Ford (2001) did, we adopted
$E_{(B-V)}=0.10$. Besides, we adopted
the extinction curve presented by Zombeck (1990).
An extinction correction $A_{\lambda}=R_{\lambda}E(B-V)$ was
applied; here $R_{\lambda}$ is obtained by interpolating
using the data of Zombeck (1990).

Since we model the stellar populations of the star clusters
by SSPs, the intrinsic colors for
each star cluster are determined by two parameters: age, and metallicity.
We will determine the ages and best-fitted models
of metallicity for these
star clusters simultaneously by a least-squares method.
The age and best-fitted model of metallicity
are found by minimizing the difference between the intrinsic and
integrated colors of GSSP:

\beq 
R^2(n,t,Z)=\sum_{i=3}^{15}[C_{\lambda_i}^{\rm
intr}(n)-C_{\lambda_i}^ {\rm ssp}(t, Z)]^2, 
\eeq
where $C_{\lambda_i}^{\rm ssp}(t, Z)$ represents the integrated color in
the $i$th filter of a SSP with age $t$ and metallicity $Z$,
and $C_{\lambda_i}^{\rm intr}(n)$ is the intrinsic
integrated color for nth star cluster.
Using the stellar evolutionary models (Bertelli et al. 1994)
and published line indices of 22 M33 older clusters,
Chandar, Bianchi, \&
Ford (1999b) narrowed the range of cluster metallicities (Z)
to be from $\sim 0.0002$ to 0.03.  So,
we only select the models of three metallicities, 0.0004, 0.004 and 0.02 of
GSSP.

Figure 3 shows the map of the best fit integrated SSP colors
(thick line) to the intrinsic integrated colors (filled circles)
for 37 globular cluster candidates.
Table 4 presents the best-fitted models of metallicities and ages for
37 globular cluster candidates. We also list the age estimates
for the other 14 cluster candidates in Table 4. The details
about these candidates can be found in Me et al. (2001, 2002a, 2002b).
From Figure 3, we can see that clusters 23 and 30
have strong emission lines. In the process of fitting, we did
not use the strong emission lines.

\begin{figure}
\figurenum{3}
\centerline{\psfig{file=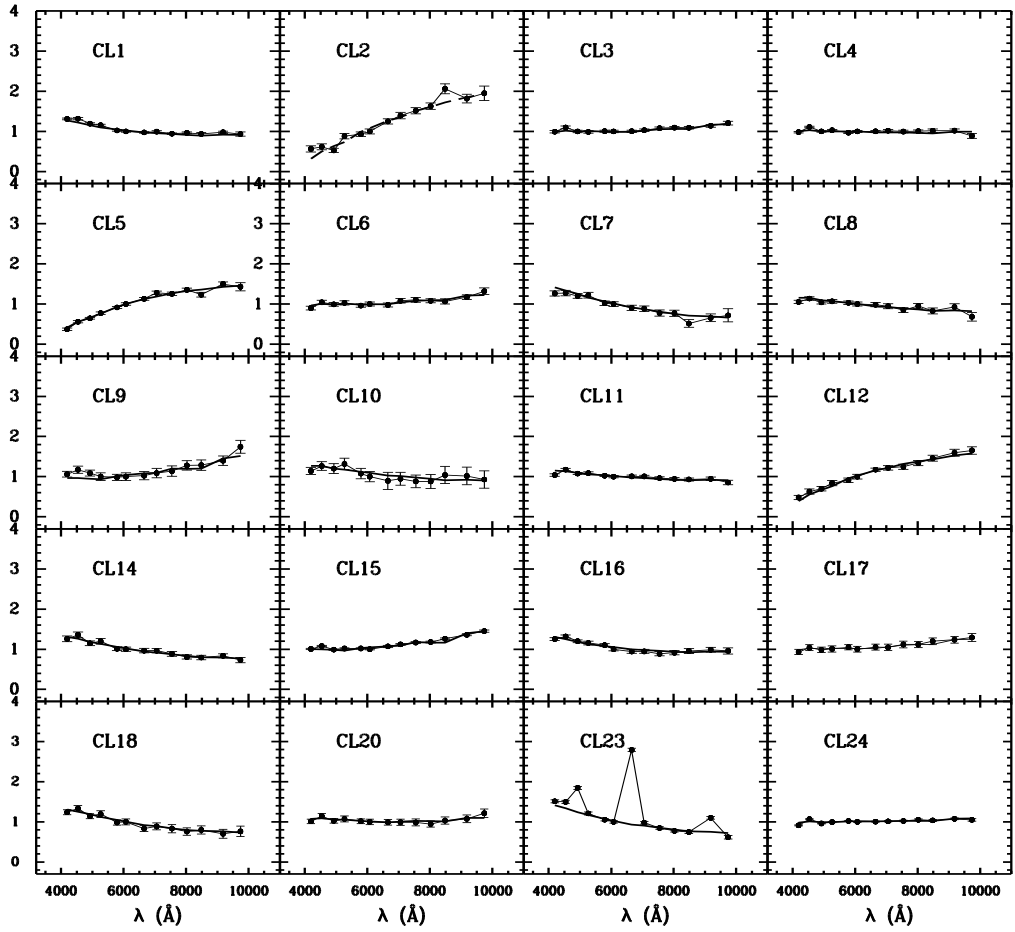,width=18.0cm,angle=270}}
\caption{Map of the best fit of the integrated color
of a SSP with intrinsic integrated color for 37 globular clusters.
Thick line represents the integrated color of a SSP, and
filled circle represents the intrinsic integrated color of a star cluster.}
\end{figure}

\begin{figure}
\figurenum{3}
\centerline{\psfig{file=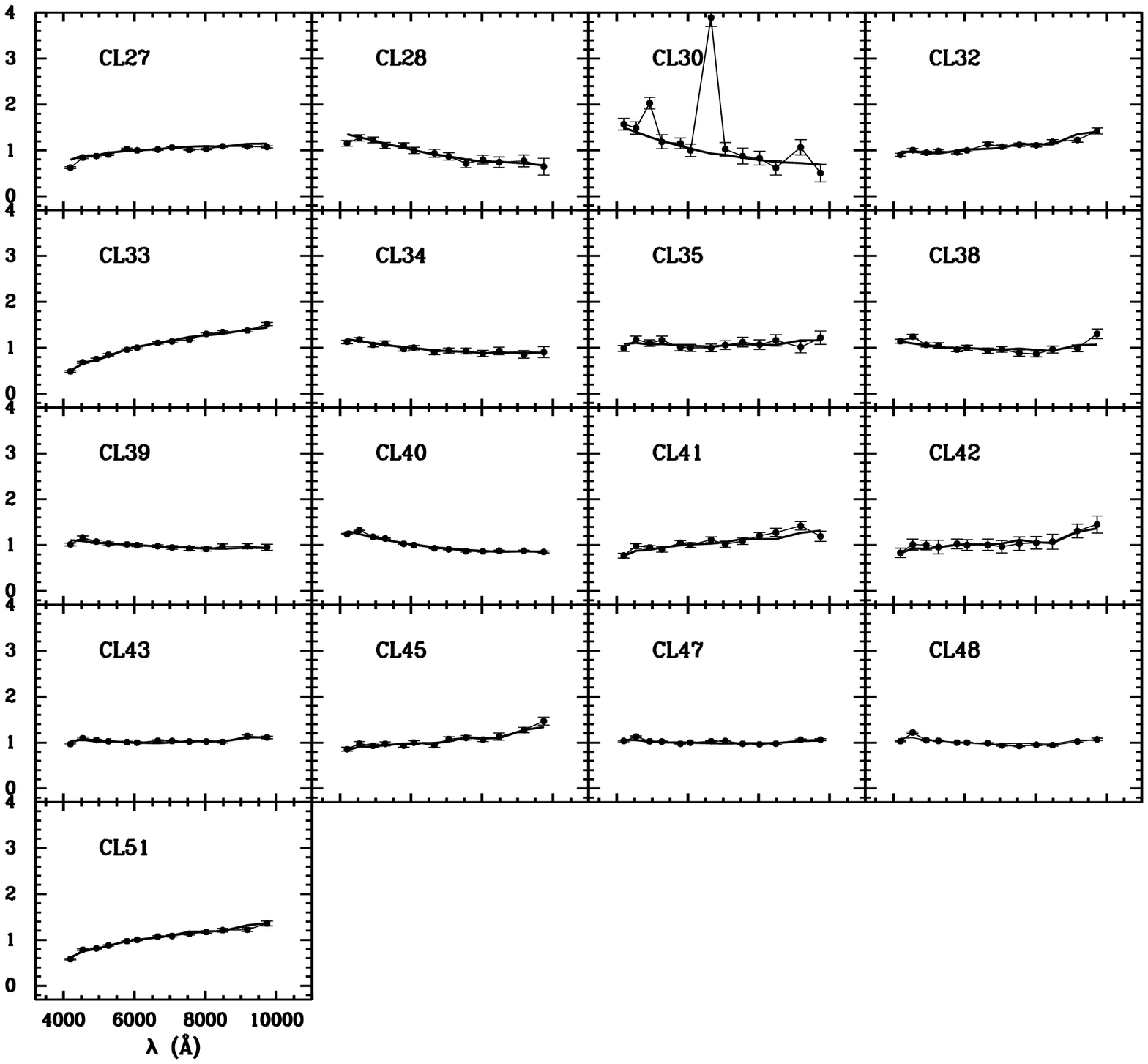,width=18.0cm,angle=270}}
\caption{Continued}
\end{figure}

Figure 4 presents a histogram of ages for the 51 globular cluster candidates.
The results show that,
in general, M33 globular clusters have been forming continuously, with
ages of $\sim 4 \times 10^{6}$ -- $10^{10}$ years, and half candidates
are younger than $10^{8}$.
There exist three groups of clusters that formed with three
metallicities, $Z=0.02, 0.004$, and $0.0004$. For different metallicities,
the distribution of cluster ages is a little different, too.
For $Z=0.02$, the ages of most clusters
are younger than $\sim 10^{7}$ years.
For $Z=0.004, 0.0004$,
the clusters formed from $\sim 4 \times 10^{6}$ -- $10^{10}$ years.

In this section, we estimate
the ages of 37 globular cluster candidates in M33
by comparing the photometry
of each object with the theoretical stellar population
synthesis models for different values of metallicity.
However, for clusters older than several $10^8$ years,
the age/metallicity degeneracy becomes
pronounced. In this case, we only mean that for some
metallicity, the intrinsic integrated color of a cluster
provides the best fit to the integrated color of a SSP at some age.
The uncertainties in the age estimated arising from photometric
uncertainties have typical values 0.2 (in $\log$ years).

\subsection{Comparison with Previous Results}

By comparing the photometric measurements to integrated colors
from theoretical models by Bertelli et al. (1994), Chandar et al.
(1999b, 2002) estimated ages for 35 globular cluster
candidates in common. Figure 5 plots the comparison of distribution of age with
previous results (Chandar et al. 1999b, 2002). Table 6 
lists this comparison. In order to test whether our results
are consistent with Chandar et al. (1999b, 2002), we
provide the Kolmogorov-Smirnov Test. The results are that
the maximum value of the absolute difference is 0.48, and
the significance level probability is 100.00 per cent.

\begin{figure}
\figurenum{4}
\centerline{\psfig{file=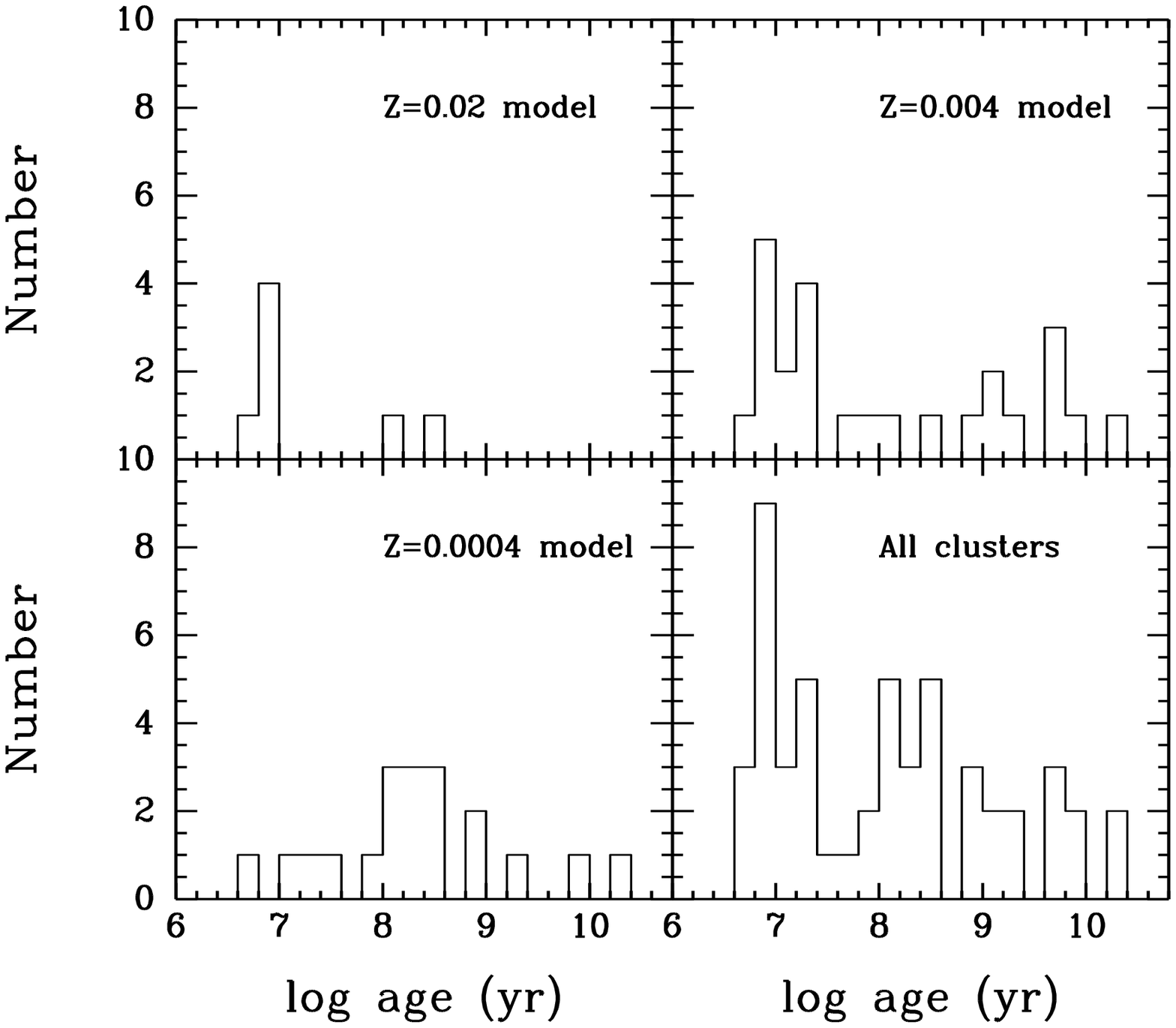,width=14.0cm,angle=-90}}
\vspace{-1.0cm}
\caption{Histogram of ages for 51 globular cluster candidates}
\label{fig4}
\end{figure}

\begin{figure}
\figurenum{5}
\centerline{\psfig{file=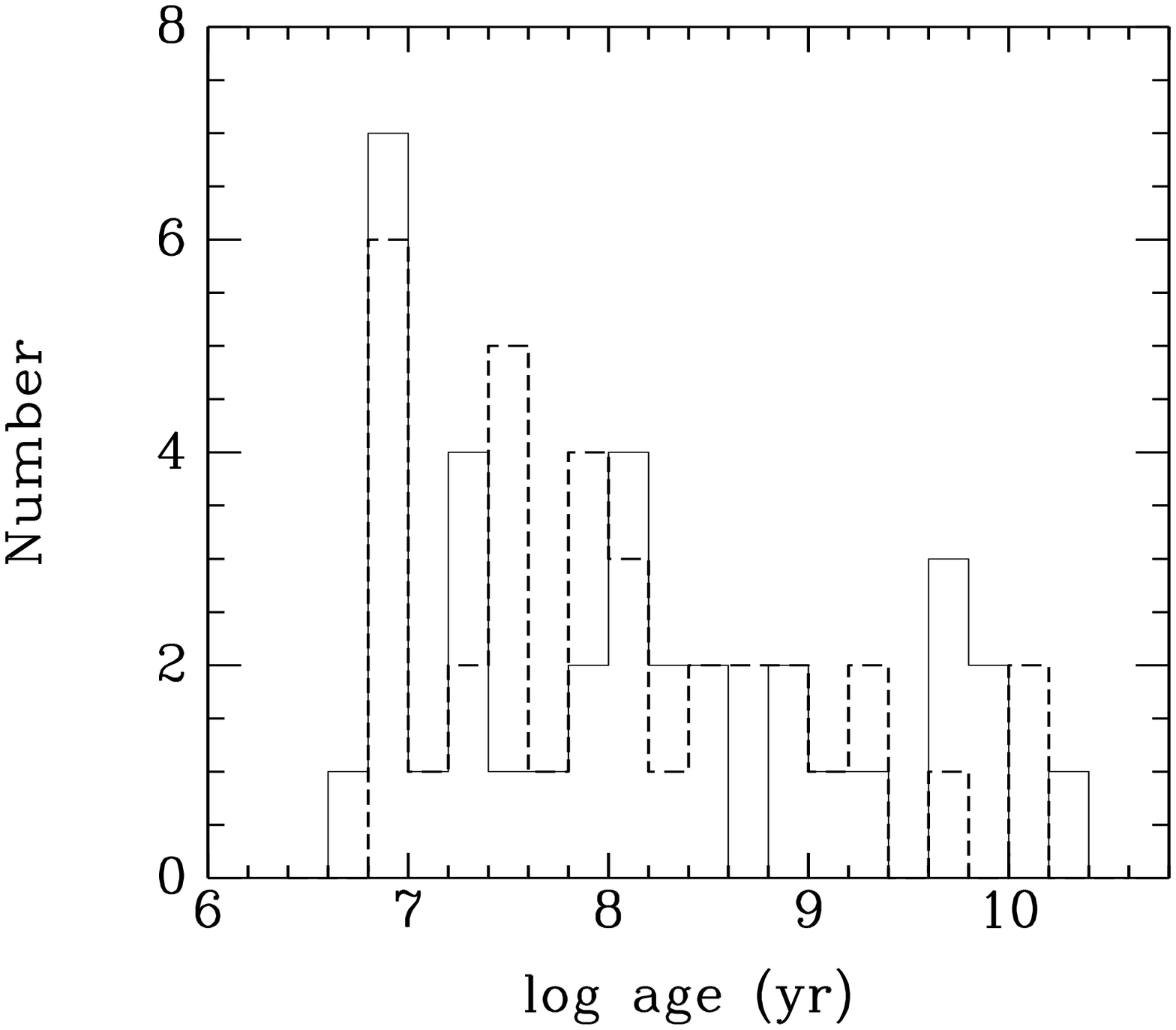,width=14.0cm,angle=-90}}
\caption{Comparison of distribution of age. The solid
histogram gives the age estimates in this paper and the
dashed histogram shows the results of Chandar et al. (1999b, 2002).}
\label{fig4}
\end{figure}

\section{SUMMARY AND DISCUSSION}

In this paper, we have, for the first time, obtained the SEDs
of 37 globular cluster candidates of M33 in 13 intermediate colors with
the BAO 60/90 cm Schmidt telescope.
Below, we summarize our main conclusions.

1. Using the images obtained with the Beijing Astronomical
Observatory $60/90$ cm Schmidt Telescope in 13 intermediate-band filters
from 3800 to 10000{\AA}, we obtained the spectral energy
distributions (SEDs) of 37 globular cluster candidates
that were detected by Mochejska et al. (1998).

2. By comparing the integrated photometric measurements with
theoretical stellar population synthesis models,
we find that clusters formed continuously in
M33 from $\sim 4\times10^6$ -- $10^{10}$ years.
The results also show that, half of the candidates are younger than
$10^8$ years.

As we know, integrated colors of star clusters depend mostly on age, with
a secondary dependence on chemical composition.
So, we can estimate ages of clusters, but cannot determine
metallicities of clusters with precision. As Chandar, Bianchi, \& Ford (1999b, 1999c)
did, we also estimated the ages of our sample clusters by comparing
the photometry of each object with models for different values of metallicity.
Although we presented the metallicity of each cluster in Table 4,
we only mean that, for this metallicity, the intrinsic integrated color
of each cluster provides the best fit to the integrated color of a SSP.

\acknowledgments
We would like to thank the anonymous referee for his/her
insightful comments and suggestions that improved this paper.
We are grateful to the Padova group for providing us with a set of
theoretical isochrones and SSPs. We also thank G. Bruzual and
S. Charlot for sending us their latest calculations of SSPs and
for explanations of their code. The BATC Survey is supported by the
Chinese Academy of Sciences, the Chinese National Natural Science
Foundation and the Chinese State Committee of Sciences and
Technology.
The project is also supported in part
by the National Science Foundation (grant INT 93-01805) and
by Arizona State University, the University of Arizona and Western
Connecticut State University.


\clearpage
\input{majuntable1.tex}
\clearpage
\input{majuntable2.tex}
\clearpage
\input{majuntable3a.tex}
\clearpage
\input{majuntable3b.tex}
\clearpage
\input{majuntable4.tex}
\clearpage
\input{majuntable5.tex}

\clearpage
\input{majuntable6.tex}

\end{document}

%% file: majuntable1.tex
\setcounter{table}{0}
\begin{table}[htb]
\caption{Parameters of the BATC filters and statistics of observations}
\vspace {0.3cm}
\begin{tabular}{cccccc}
\hline
\hline
 No. & Name& cw\tablenotemark{a}~~(\AA)& Exp. (hr)&  N.img\tablenotemark{b}
 & rms\tablenotemark{c} \\
\hline
1  & BATC03& 4210   & 00:55& 04 &0.024\\
2  & BATC04& 4546   & 01:05& 04 &0.023\\
3  & BATC05& 4872   & 03:55& 19 &0.017\\
4  & BATC06& 5250   & 03:19& 15 &0.006\\
5  & BATC07& 5785   & 04:38& 17 &0.011\\
6  & BATC08& 6075   & 01:26& 08 &0.016\\
7  & BATC09& 6710   & 01:09& 08 &0.006\\
8  & BATC10& 7010   & 01:41& 08 &0.005\\
9  & BATC11& 7530   & 02:07& 10 &0.017\\
10 & BATC12& 8000   & 03:00& 11 &0.003\\
11 & BATC13& 8510   & 03:15& 11 &0.005\\
12 & BATC14& 9170   & 05:45& 25 &0.011\\
13 & BATC15& 9720   & 06:00& 26 &0.009\\
\hline
\end{tabular}\\
\tablenotetext{a}{Central wavelength for each BATC filter}
\tablenotetext{b}{Image numbers for each BATC filter}
\tablenotetext{c}{Zero point error, in magnitude, for each filter
as obtained from the standard stars}
\end{table}

%% file: majuntable2.tex
\setcounter{table}{1}
\begin{table}[htb]
\caption{Comparison of common clusters in different studies}
\vspace{0.3cm}
\begin{tabular}{cccc}
\hline
\hline
MKSS98 & MD78 & CS82 & CBF99 \& CBF01 \\
\hline
4      & 10   & .... & ....           \\
11     & .... & U11  & ....           \\
13     & .... & .... & 44*            \\
19     & .... & .... & 47*            \\
20     & .... & U94  & ....           \\
21     & 24   & U49* & 61             \\
22     & 25   & H38* & 104            \\
24     & 27   & .... & ....           \\
25     & .... & .... & 49*            \\
26     & .... & H14  & 33*            \\
28     & .... & U75  & ....           \\
29     & .... & .... & 28*            \\
31     & .... & R14* & 98             \\
36     & .... & R12* & 116            \\
37     & .... & .... & 141*           \\
38     & 35   & H30  & ....           \\
39     & 39   & U62  & ....           \\
41     & 38   & U83  & ....           \\
42     & 41   & U78  & ....           \\
44     & 42   & U82  & 55*            \\
45     & 36   & H33  & ....           \\
46     & .... & U79  & 155*           \\
49     & .... & .... & 35*            \\
50     & .... & U91  & 58*            \\
51     & 45   & H21  & ....           \\
\hline
\end{tabular}
\end{table}

%% file: majuntable3a.tex
{\small
\setcounter{table}{2}
\begin{table}[htb]
\caption{SEDs of 37 globular cluster candidates}
\vspace {0.3cm}
\begin{tabular}{cccccccccccccc}
\hline
\hline
 No. & 03  &  04 &  05 &  06 &  07 &  08 &  09 &  10 &  11 &  12 &  13 &  14 &  15\\
(1)    & (2) & (3) & (4) & (5) & (6) & (7) & (8) & (9) & (10) & (11) & (12) & (13) & (14)\\
\hline
       1 & 17.153 & 17.116 & 17.194 & 17.191 & 17.271 & 17.287 & 17.284 & 17.251 & 17.290 & 17.247 & 17.258 & 17.190 & 17.232\\
         & 0.019 &  0.021 &  0.019 &  0.024 &  0.020 &  0.022 &  0.019 &  0.025 &  0.024 &  0.024 &  0.043 &  0.032 &  0.061 \\
       2 & 19.660 & 19.533 & 19.676 & 19.079 & 18.966 & 18.881 & 18.610 & 18.480 & 18.369 & 18.272 & 17.995 & 18.111 & 18.024\\
         & 0.151 &  0.129 &  0.112 &  0.083 &  0.064 &  0.066 &  0.055 &  0.062 &  0.055 &  0.054 &  0.065 &  0.065 &  0.099 \\
       3 & 17.108 & 16.968 & 17.042 & 17.014 & 16.945 & 16.941 & 16.906 & 16.868 & 16.797 & 16.761 & 16.745 & 16.677 & 16.606\\
         & 0.036 &  0.034 &  0.031 &  0.034 &  0.027 &  0.030 &  0.031 &  0.031 &  0.031 &  0.030 &  0.033 &  0.032 &  0.042 \\
       4 & 17.476 & 17.313 & 17.396 & 17.326 & 17.340 & 17.297 & 17.265 & 17.235 & 17.240 & 17.211 & 17.179 & 17.154 & 17.295\\
         & 0.022 &  0.021 &  0.021 &  0.024 &  0.022 &  0.024 &  0.027 &  0.031 &  0.038 &  0.037 &  0.047 &  0.045 &  0.074 \\
       5 & 19.174 & 18.700 & 18.515 & 18.279 & 18.046 & 17.939 & 17.776 & 17.631 & 17.633 & 17.530 & 17.612 & 17.380 & 17.421\\
         & 0.084 &  0.051 &  0.037 &  0.043 &  0.031 &  0.034 &  0.032 &  0.036 &  0.037 &  0.035 &  0.050 &  0.038 &  0.076 \\
       6 & 17.593 & 17.404 & 17.429 & 17.346 & 17.371 & 17.319 & 17.313 & 17.191 & 17.152 & 17.153 & 17.152 & 17.027 & 16.888\\
         & 0.055 &  0.053 &  0.050 &  0.055 &  0.046 &  0.054 &  0.055 &  0.056 &  0.056 &  0.050 &  0.066 &  0.055 &  0.069 \\
       7 & 18.219 & 18.181 & 18.214 & 18.158 & 18.299 & 18.313 & 18.390 & 18.405 & 18.528 & 18.517 & 18.933 & 18.651 & 18.542\\
         & 0.054 &  0.058 &  0.053 &  0.064 &  0.058 &  0.065 &  0.075 &  0.082 &  0.106 &  0.108 &  0.205 &  0.159 &  0.248 \\
       8 & 18.030 & 17.916 & 17.981 & 17.916 & 17.901 & 17.925 & 17.911 & 17.946 & 18.041 & 17.901 & 18.028 & 17.888 & 18.207\\
         & 0.042 &  0.042 &  0.041 &  0.050 &  0.046 &  0.054 &  0.057 &  0.070 &  0.077 &  0.072 &  0.099 &  0.095 &  0.168 \\
       9 & 17.865 & 17.727 & 17.779 & 17.829 & 17.799 & 17.769 & 17.713 & 17.638 & 17.568 & 17.422 & 17.396 & 17.287 & 17.036\\
         & 0.072 &  0.084 &  0.074 &  0.106 &  0.084 &  0.106 &  0.106 &  0.117 &  0.124 &  0.101 &  0.108 &  0.091 &  0.101 \\
      10 & 18.580 & 18.439 & 18.465 & 18.325 & 18.494 & 18.563 & 18.654 & 18.581 & 18.641 & 18.623 & 18.420 & 18.426 & 18.514\\
         & 0.088 &  0.093 &  0.110 &  0.116 &  0.132 &  0.136 &  0.256 &  0.184 &  0.196 &  0.212 &  0.222 &  0.233 &  0.251 \\
      11 & 16.928 & 16.765 & 16.831 & 16.780 & 16.795 & 16.811 & 16.775 & 16.766 & 16.792 & 16.795 & 16.790 & 16.752 & 16.854\\
         & 0.031 &  0.029 &  0.027 &  0.031 &  0.025 &  0.029 &  0.034 &  0.033 &  0.038 &  0.040 &  0.043 &  0.048 &  0.070 \\
      12 & 17.993 & 17.672 & 17.525 & 17.289 & 17.151 & 17.037 & 16.833 & 16.781 & 16.737 & 16.640 & 16.527 & 16.410 & 16.361\\
         & 0.116 &  0.114 &  0.089 &  0.086 &  0.064 &  0.064 &  0.038 &  0.051 &  0.055 &  0.051 &  0.056 &  0.058 &  0.061 \\
      14 & 16.305 & 16.192 & 16.345 & 16.258 & 16.391 & 16.394 & 16.404 & 16.394 & 16.468 & 16.548 & 16.546 & 16.467 & 16.604\\
         & 0.060 &  0.055 &  0.054 &  0.063 &  0.050 &  0.056 &  0.062 &  0.061 &  0.074 &  0.071 &  0.079 &  0.066 &  0.097 \\
      15 & 15.844 & 15.739 & 15.802 & 15.732 & 15.681 & 15.691 & 15.588 & 15.529 & 15.463 & 15.432 & 15.339 & 15.238 & 15.157\\
         & 0.018 &  0.015 &  0.018 &  0.016 &  0.015 &  0.015 &  0.027 &  0.019 &  0.020 &  0.019 &  0.020 &  0.019 &  0.020 \\
      16 & 17.061 & 16.975 & 17.049 & 17.050 & 17.052 & 17.146 & 17.180 & 17.167 & 17.216 & 17.164 & 17.096 & 17.040 & 17.063\\
         & 0.030 &  0.031 &  0.030 &  0.041 &  0.035 &  0.044 &  0.051 &  0.056 &  0.065 &  0.061 &  0.070 &  0.065 &  0.086 \\
      17 & 16.717 & 16.569 & 16.596 & 16.531 & 16.438 & 16.480 & 16.395 & 16.391 & 16.299 & 16.279 & 16.183 & 16.129 & 16.072\\
         & 0.077 &  0.078 &  0.075 &  0.082 &  0.067 &  0.077 &  0.076 &  0.082 &  0.082 &  0.076 &  0.075 &  0.074 &  0.084 \\
      18 & 16.962 & 16.854 & 16.991 & 16.903 & 17.063 & 17.035 & 17.204 & 17.128 & 17.177 & 17.255 & 17.182 & 17.298 & 17.202\\
         & 0.055 &  0.058 &  0.060 &  0.074 &  0.070 &  0.080 &  0.103 &  0.106 &  0.133 &  0.132 &  0.140 &  0.170 &  0.185 \\
      20 & 17.189 & 17.027 & 17.114 & 17.024 & 17.036 & 17.042 & 17.024 & 17.014 & 16.999 & 17.023 & 16.910 & 16.843 & 16.702\\
         & 0.059 &  0.060 &  0.060 &  0.075 &  0.064 &  0.074 &  0.079 &  0.085 &  0.095 &  0.090 &  0.095 &  0.095 &  0.095 \\
      23 & 16.636 & 16.617 & 16.358 & 16.778 & 16.877 & 16.924 & 15.780 & 16.909 & 17.047 & 17.124 & 17.143 & 16.701 & 17.321\\
         & 0.026 &  0.028 &  0.022 &  0.034 &  0.029 &  0.033 &  0.016 &  0.035 &  0.042 &  0.041 &  0.058 &  0.035 &  0.076 \\
      24 & 16.764 & 16.563 & 16.649 & 16.570 & 16.490 & 16.503 & 16.468 & 16.450 & 16.420 & 16.369 & 16.366 & 16.304 & 16.320\\
         & 0.015 &  0.013 &  0.013 &  0.014 &  0.013 &  0.014 &  0.015 &  0.016 &  0.020 &  0.020 &  0.023 &  0.024 &  0.032 \\
      27 & 16.437 & 16.080 & 16.005 & 15.929 & 15.736 & 15.763 & 15.710 & 15.657 & 15.688 & 15.652 & 15.570 & 15.554 & 15.553\\
         & 0.035 &  0.027 &  0.022 &  0.021 &  0.015 &  0.018 &  0.020 &  0.018 &  0.018 &  0.016 &  0.018 &  0.020 &  0.022 \\
      28 & 17.920 & 17.784 & 17.795 & 17.867 & 17.818 & 17.916 & 17.956 & 18.028 & 18.215 & 18.079 & 18.138 & 18.074 & 18.264\\
         & 0.058 &  0.058 &  0.055 &  0.064 &  0.061 &  0.071 &  0.102 &  0.099 &  0.142 &  0.130 &  0.167 &  0.182 &  0.311 \\
      30 & 17.084 & 17.114 & 16.746 & 17.289 & 17.269 & 17.415 & 15.908 & 17.348 & 17.496 & 17.539 & 17.834 & 17.224 & 18.026\\
         & 0.089 &  0.101 &  0.068 &  0.139 &  0.110 &  0.146 &  0.056 &  0.160 &  0.215 &  0.196 &  0.280 &  0.168 &  0.415 \\
      32 & 17.507 & 17.360 & 17.394 & 17.314 & 17.298 & 17.239 & 17.070 & 17.117 & 17.053 & 17.043 & 16.953 & 16.898 & 16.726\\
         & 0.043 &  0.037 &  0.035 &  0.038 &  0.033 &  0.033 &  0.039 &  0.038 &  0.038 &  0.039 &  0.044 &  0.039 &  0.051 \\
      33 & 17.060 & 16.652 & 16.515 & 16.349 & 16.163 & 16.107 & 15.968 & 15.927 & 15.868 & 15.741 & 15.686 & 15.640 & 15.525\\
         & 0.045 &  0.034 &  0.026 &  0.031 &  0.022 &  0.024 &  0.022 &  0.022 &  0.023 &  0.020 &  0.022 &  0.021 &  0.024 \\
      34 & 17.545 & 17.465 & 17.550 & 17.474 & 17.558 & 17.516 & 17.597 & 17.541 & 17.535 & 17.578 & 17.495 & 17.567 & 17.495\\
         & 0.037 &  0.041 &  0.042 &  0.048 &  0.048 &  0.052 &  0.071 &  0.064 &  0.078 &  0.084 &  0.094 &  0.105 &  0.143 \\
      35 & 17.830 & 17.608 & 17.652 & 17.547 & 17.660 & 17.657 & 17.630 & 17.554 & 17.470 & 17.505 & 17.394 & 17.521 & 17.311\\
         & 0.071 &  0.068 &  0.069 &  0.079 &  0.077 &  0.085 &  0.093 &  0.097 &  0.102 &  0.106 &  0.113 &  0.130 &  0.129 \\
\end{tabular}
\end{table}

%% file: majuntable3b.tex
{\small
\setcounter{table}{2}
\begin{table}[htb]
\caption{Continued}
\vspace {0.3cm}
\begin{tabular}{cccccccccccccc}
\hline
\hline
 No. & 03  &  04 &  05 &  06 &  07 &  08 &  09 &  10 &  11 &  12 &  13 &  14 &  15\\
(1)    & (2) & (3) & (4) & (5) & (6) & (7) & (8) & (9) & (10) & (11) & (12) & (13) & (14)\\
\hline
      38 & 17.551 & 17.430 & 17.570 & 17.543 & 17.591 & 17.536 & 17.580 & 17.534 & 17.607 & 17.614 & 17.476 & 17.427 & 17.118\\
         & 0.042 &  0.043 &  0.046 &  0.064 &  0.051 &  0.060 &  0.069 &  0.070 &  0.083 &  0.080 &  0.087 &  0.083 &  0.086 \\
      39 & 17.649 & 17.469 & 17.526 & 17.532 & 17.498 & 17.505 & 17.500 & 17.519 & 17.524 & 17.519 & 17.437 & 17.410 & 17.425\\
         & 0.035 &  0.035 &  0.033 &  0.043 &  0.032 &  0.037 &  0.037 &  0.045 &  0.052 &  0.051 &  0.066 &  0.060 &  0.077 \\
      40 & 15.796 & 15.689 & 15.789 & 15.784 & 15.848 & 15.870 & 15.916 & 15.934 & 15.967 & 15.950 & 15.913 & 15.895 & 15.915\\
         & 0.011 &  0.012 &  0.011 &  0.013 &  0.011 &  0.014 &  0.014 &  0.016 &  0.018 &  0.017 &  0.019 &  0.020 &  0.028 \\
      41 & 18.172 & 17.882 & 17.887 & 17.894 & 17.686 & 17.731 & 17.577 & 17.669 & 17.577 & 17.448 & 17.367 & 17.229 & 17.409\\
         & 0.072 &  0.056 &  0.049 &  0.070 &  0.049 &  0.057 &  0.060 &  0.071 &  0.074 &  0.061 &  0.077 &  0.070 &  0.103 \\
      42 & 18.444 & 18.204 & 18.182 & 18.195 & 18.069 & 18.087 & 18.051 & 18.082 & 17.994 & 17.951 & 17.910 & 17.675 & 17.553\\
         & 0.134 &  0.132 &  0.114 &  0.168 &  0.112 &  0.128 &  0.134 &  0.152 &  0.157 &  0.141 &  0.163 &  0.126 &  0.142 \\
      43 & 16.576 & 16.408 & 16.420 & 16.409 & 16.375 & 16.377 & 16.306 & 16.304 & 16.292 & 16.272 & 16.261 & 16.114 & 16.132\\
         & 0.019 &  0.019 &  0.018 &  0.018 &  0.015 &  0.017 &  0.021 &  0.018 &  0.018 &  0.018 &  0.025 &  0.019 &  0.028 \\
      45 & 18.239 & 18.061 & 18.089 & 18.000 & 17.994 & 17.909 & 17.946 & 17.787 & 17.741 & 17.761 & 17.677 & 17.527 & 17.364\\
         & 0.054 &  0.047 &  0.045 &  0.047 &  0.041 &  0.043 &  0.057 &  0.047 &  0.045 &  0.045 &  0.073 &  0.051 &  0.066 \\
      47 & 16.659 & 16.531 & 16.605 & 16.568 & 16.570 & 16.533 & 16.473 & 16.458 & 16.503 & 16.495 & 16.459 & 16.353 & 16.337\\
         & 0.015 &  0.014 &  0.012 &  0.014 &  0.012 &  0.013 &  0.011 &  0.015 &  0.015 &  0.015 &  0.023 &  0.021 &  0.021 \\
      48 & 16.752 & 16.540 & 16.667 & 16.643 & 16.631 & 16.622 & 16.611 & 16.656 & 16.649 & 16.598 & 16.585 & 16.476 & 16.418\\
         & 0.016 &  0.014 &  0.012 &  0.014 &  0.012 &  0.012 &  0.013 &  0.015 &  0.016 &  0.017 &  0.022 &  0.023 &  0.025 \\
      51 & 18.066 & 17.704 & 17.638 & 17.515 & 17.354 & 17.315 & 17.210 & 17.183 & 17.119 & 17.062 & 17.006 & 16.977 & 16.851\\
         & 0.030 &  0.022 &  0.018 &  0.021 &  0.017 &  0.019 &  0.019 &  0.022 &  0.023 &  0.024 &  0.038 &  0.035 &  0.042 \\
\hline
\end{tabular}
\end{table}

%% file: majuntable4.tex
\setcounter{table}{3}
\begin{table}[htb]
\caption{Comparison of cluster photometry with previous measurements}
\vspace {0.3cm}
\begin{tabular}{ccccc}
\hline
\hline
No. & $V$ (MKKSS98)  & $V$ (BATC) &$B-V$ (MKKSS98) &$B-V$ (BATC) \\
(1)    & (2) & (3) & (4) & (5) \\
\hline
      1...... &  17.54 & 17.299 $\pm$  0.035 &   0.06 &  -0.118 $\pm$  0.046\\
      2...... &  19.69 & 19.089 $\pm$  0.112 &   2.50 &   0.515 $\pm$  0.218\\
      3...... &  17.08 & 17.028 $\pm$  0.048 &   0.22 &   0.029 $\pm$  0.068\\
      4...... &  17.29 & 17.408 $\pm$  0.038 &   0.25 &  -0.004 $\pm$  0.048\\
      5...... &  18.34 & 18.215 $\pm$  0.056 &   0.73 &   0.705 $\pm$  0.096\\
      6...... &  17.78 & 17.439 $\pm$  0.081 &   0.26 &   0.076 $\pm$  0.111\\
      7...... &  18.36 & 18.308 $\pm$  0.100 &   0.17 &  -0.052 $\pm$  0.129\\
      8...... &  18.07 & 17.957 $\pm$  0.080 &   0.18 &   0.044 $\pm$  0.100\\
      9...... &  18.34 & 17.877 $\pm$  0.153 &   0.07 &  -0.057 $\pm$  0.192\\
     10...... &  18.49 & 18.476 $\pm$  0.213 &   0.03 &   0.063 $\pm$  0.254\\
     11...... &  16.97 & 16.844 $\pm$  0.044 &   0.21 &   0.017 $\pm$  0.061\\
     12...... &  17.40 & 17.291 $\pm$  0.112 &   0.66 &   0.558 $\pm$  0.195\\
     13...... &  17.62 & 17.661 $\pm$  0.052 &   0.27 &  -0.030 $\pm$  0.093\\
     14...... &  16.35 & 16.406 $\pm$  0.088 &   0.14 &  -0.149 $\pm$  0.119\\
     15...... &  15.81 & 15.753 $\pm$  0.025 &   0.28 &   0.069 $\pm$  0.034\\
     16...... &  17.11 & 17.080 $\pm$  0.062 &   0.11 &  -0.028 $\pm$  0.077\\
     17...... &  16.99 & 16.513 $\pm$  0.118 &   0.35 &   0.156 $\pm$  0.163\\
     18...... &  17.07 & 17.079 $\pm$  0.120 &   0.05 &  -0.158 $\pm$  0.146\\
     19...... &  17.75 & 17.323 $\pm$  0.040 &  -0.21 &   0.137 $\pm$  0.072\\
     20...... &  17.43 & 17.089 $\pm$  0.112 &  -0.12 &   0.029 $\pm$  0.142\\
     21...... &  16.04 & 16.207 $\pm$  0.019 &   0.78 &   0.549 $\pm$  0.032\\
     22...... &  17.26 & 17.208 $\pm$  0.028 &   0.81 &   0.664 $\pm$  0.049\\
     23...... &  17.18 & 16.889 $\pm$  0.051 &  -0.03 &  -0.136 $\pm$  0.064\\
     24...... &  16.53 & 16.571 $\pm$  0.022 &   0.29 &   0.092 $\pm$  0.029\\
     25...... &  18.21 & 18.285 $\pm$  0.092 &   0.44 &   0.507 $\pm$  0.213\\
     26...... &  17.05 & 17.141 $\pm$  0.031 &   0.26 &   0.118 $\pm$  0.060\\
     27...... &  16.05 & 15.849 $\pm$  0.028 &   0.61 &   0.401 $\pm$  0.048\\
     28...... &  17.88 & 17.861 $\pm$  0.105 &   0.13 &   0.025 $\pm$  0.134\\
     29...... &  16.23 & 16.322 $\pm$  0.019 &   0.82 &   0.652 $\pm$  0.044\\
     30...... &  17.05 & 17.287 $\pm$  0.202 &   0.06 &  -0.024 $\pm$  0.244\\
     31...... &  16.49 & 16.414 $\pm$  0.035 &   0.95 &   0.857 $\pm$  0.065\\
     32...... &  17.59 & 17.381 $\pm$  0.056 &   0.27 &   0.078 $\pm$  0.078\\
     33...... &  16.31 & 16.300 $\pm$  0.040 &   0.71 &   0.547 $\pm$  0.064\\ 
     34...... &  17.88 & 17.603 $\pm$  0.080 &   0.07 &  -0.065 $\pm$  0.099\\
     35...... &  17.98 & 17.683 $\pm$  0.130 &  -0.01 &   0.038 $\pm$  0.163\\
     36...... &  16.32 & 16.310 $\pm$  0.036 &   0.90 &   0.736 $\pm$  0.064\\
     37...... &  16.17 & 16.250 $\pm$  0.074 &   0.08 &  -0.202 $\pm$  0.095\\
     38...... &  17.76 & 17.652 $\pm$  0.091 &   0.17 &  -0.152 $\pm$  0.110\\
     39...... &  17.39 & 17.566 $\pm$  0.058 &   0.26 &   0.005 $\pm$  0.077\\
     40...... &  15.98 & 15.879 $\pm$  0.020 &   0.09 &  -0.115 $\pm$  0.026\\
     41...... &  17.82 & 17.798 $\pm$  0.090 &   0.43 &   0.222 $\pm$  0.122\\
     42...... &  18.16 & 18.163 $\pm$  0.208 &   0.41 &   0.173 $\pm$  0.279\\
     43...... &  16.52 & 16.444 $\pm$  0.026 &   0.29 &   0.072 $\pm$  0.038\\
     44...... &  17.90 & 18.112 $\pm$  0.068 &   0.48 &   0.295 $\pm$  0.135\\
     45...... &  18.07 & 18.082 $\pm$  0.070 &   0.32 &   0.086 $\pm$  0.098\\
     46...... &  17.49 & 17.546 $\pm$  0.055 &   0.24 &   0.066 $\pm$  0.069\\
     47...... &  16.66 & 16.640 $\pm$  0.021 &   0.25 &  -0.023 $\pm$  0.029\\
     48...... &  16.69 & 16.697 $\pm$  0.020 &   0.23 &  -0.064 $\pm$  0.029\\
     49...... &  17.08 & 17.135 $\pm$  0.031 &   0.23 &   0.012 $\pm$  0.056\\
     50...... &  18.61 & 18.468 $\pm$  0.101 &   0.30 &   0.182 $\pm$  0.190\\
     51...... &  17.37 & 17.478 $\pm$  0.030 &   0.64 &   0.395 $\pm$  0.044\\
\hline
\end{tabular}
\end{table}

%% file: majuntable5.tex
\begin{table}[htb]
\caption[]{Age distribution of 51 globular cluster candidates}
\vspace {0.3cm}
\begin{tabular}{ccc|ccc}
\hline
\hline
 No. & Metallicity ($Z$)& Age ([$\log$ yr]) & No. & Metallicity ($Z$)& Age ([$\log$ yr])\\
 (1) & (2) & (3) & (1) & (2) & (3) \\
\hline
      1...... & 0.00400 &  7.220 &	     27...... & 0.00040 &  8.957\\
      2...... & 0.00400 & 10.301 &	     28...... & 0.00400 &  6.620\\
      3...... & 0.00040 &  8.407 &	     29...... & 0.00040 &  9.796\\
      4...... & 0.00400 &  7.857 &	     30...... & 0.00040 &  7.179\\
      5...... & 0.00040 & 10.238 &	     31...... & 0.00400 &  9.110\\
      6...... & 0.00040 &  8.507 &	     32...... & 0.02000 &  7.000\\
      7...... & 0.02000 &  6.720 &	     33...... & 0.00400 &  9.207\\
      8...... & 0.00040 &  8.009 &	     34...... & 0.00400 &  7.100\\
      9...... & 0.02000 &  6.960 &	     35...... & 0.00040 &  8.307\\
     10...... & 0.00040 &  8.009 &	     36...... & 0.00400 & 10.000\\
     11...... & 0.00400 &  6.940 &	     37...... & 0.00040 &  6.660\\
     12...... & 0.00400 &  9.628 &	     38...... & 0.02000 &  6.880\\
     13...... & 0.00400 &  7.220 &	     39...... & 0.00400 &  6.960\\
     14...... & 0.00040 &  7.462 &	     40...... & 0.00400 &  7.158\\
     15...... & 0.02000 &  7.000 &	     41...... & 0.00400 &  8.806\\
     16...... & 0.00400 &  7.220 &	     42...... & 0.02000 &  8.556\\
     17...... & 0.00040 &  8.507 &	     43...... & 0.00040 &  8.307\\
     18...... & 0.00400 &  6.820 &	     44...... & 0.00040 &  8.957\\
     19...... & 0.02000 &  8.057 &	     45...... & 0.00400 &  8.556\\
     20...... & 0.00040 &  8.255 &	     46...... & 0.00400 &  6.960\\
     21...... & 0.00400 &  9.600 &	     47...... & 0.00400 &  7.380\\
     22...... & 0.00400 &  9.700 &	     48...... & 0.00040 &  8.057\\
     23...... & 0.00040 &  7.320 &	     49...... & 0.00040 &  7.806\\
     24...... & 0.00400 &  7.699 &	     50...... & 0.00400 &  6.960\\
     25...... & 0.00040 &  9.342 &	     51...... & 0.00400 &  9.009\\
     26...... & 0.00400 &  8.009 &	              &         &       \\
\hline
\end{tabular}
\end{table}

%% file: majuntable6.tex
\begin{table}[htb]
\caption[]{Comparison of age estimates for globular cluster candidates with previous measurements}
\vspace {0.3cm}
\begin{tabular}{ccc|ccc}
\hline
\hline
     & Age ([$\log$ yr]) & Age ([$\log$ yr]) &     & Age ([$\log$ yr]) & Age ([$\log$ yr])\\
 No. &   (Chandar et al.)  & (This paper)        & No. &   (Chandar et al.)  & (This paper) \\
 (1) & (2) & (3) & (1) & (2) & (3) \\
\hline
      1...... & $7.4\pm 0.2$ &  7.220 &	     26...... & $7.9\pm 0.1$ &  8.009\\
      2...... & $>9.0$       & 10.301 &	     27...... & $8.6\pm 0.2$ &  8.957\\
      3...... & $8.1\pm 0.3$ &  8.407 &	     29...... & $10.2\pm 0.4$ &  9.796\\
      4...... & $7.5\pm 0.5$ &  7.857 &	     31...... & $10.2\pm 0.2$ &  9.110\\
      7...... & $7.0\pm 0.2$ &  6.720 &	     32...... & $8.1\pm 0.3$ &  7.000\\
      8...... & $7.6\pm 0.2$ &  8.009 &	     34...... & $7.4\pm 0.2$ &  7.100\\
      9...... & $8.5\pm 0.5$ &  6.960 &	     35...... & $7.0\pm 0.2$ &  8.307\\
     11...... & $7.2\pm 0.2$ &  6.940 &	     36...... & $9.7\pm 0.3$ & 10.000\\
     12...... & $9.4\pm 0.2$ &  9.628 &	     38...... & $<7.0$       &  6.880\\
     13...... & $<7.6$       &  7.220 &	     39...... & $<7.0$       &  6.960\\
     14...... & $<7.0$       &  7.462 &	     43...... & $8.5\pm 0.2$ &  8.307\\
     16...... & $8.0\pm 0.4$ &  7.220 &	     44...... & $9.0\pm 0.3$ &  8.957\\
     18...... & $<7.0$       &  6.820 &	     45...... & $8.4\pm 0.2$ &  8.556\\
     19...... & $<7.6$       &  8.057 &	     47...... & $<7.5$       &  7.380\\
     21...... & $9.2\pm 0.1$ &  9.600 &	     48...... & $8.0\pm 0.4$ &  8.057\\
     22...... & $9.25\pm 0.15$ &  9.700 &    49...... & $7.7\pm 0.1$ &  7.806\\
     24...... & $8.1\pm 0.3$ &  7.320 &	     50...... & $8.6\pm 0.3$ &  6.960\\
     25...... & $7.9\pm 0.2$ &  9.342 &	              &         &       \\
\hline
\end{tabular}
\end{table}